\documentclass{ws-procs961x669}
\usepackage{amssymb,amsmath,latexsym}
\usepackage{graphicx}
\usepackage{hyperref}

\newcommand\rf[1]{(\ref{eq:#1})}
\newcommand\lab[1]{\label{eq:#1}}
\newcommand\nonu{\nonumber}
\newcommand\br{\begin{eqnarray}}
\newcommand\er{\end{eqnarray}}
\newcommand\be{\begin{equation}}
\newcommand\ee{\end{equation}}

\newcommand\lb{\lbrack}
\newcommand\rb{\rbrack}

\newcommand\rrangle{\right\rangle}

\newcommand\llb{\left\lbrack}
\newcommand\rrb{\right\rbrack}

\renewcommand\({\left(}
\renewcommand\){\right)}
\newcommand\bgv{\bigg\vert}              

\newcommand\bc{\begin{center}}
\newcommand\ec{\end{center}}

\newcommand\partder[2]{\frac{{\partial {#1}}}{{\partial {#2}}}}

\renewcommand\a{\alpha}

\renewcommand\d{\delta}

\newcommand\eps{\epsilon}
\newcommand\vareps{\varepsilon}

\newcommand\G{\Gamma}

\newcommand\h{\frac{1}{2}}
\renewcommand\k{\kappa}
\renewcommand\l{\lambda}
\renewcommand\L{\Lambda}
\newcommand\m{\mu}
\newcommand\n{\nu}
\newcommand\om{\omega}
\renewcommand\O{\Omega}

\newcommand\vp{\varphi}
\renewcommand\P{\Phi}
\newcommand\pa{\partial}

\newcommand\pr{\prime}

\newcommand\s{\sigma}

\renewcommand\t{\tau}
\renewcommand\th{\theta}

\newcommand\wti{\widetilde}

\newcommand\cA{{\mathcal A}}
\newcommand\cB{{\mathcal B}}

\newcommand\cF{{\mathcal F}}

\newcommand\cH{{\mathcal H}}

\newcommand\cL{{\mathcal L}}
\newcommand\cM{{\mathcal M}}

\newcommand\cU{{\mathcal U}}

\newcommand{\ct}[1]{\cite{#1}}
\newcommand\PRL[3]{\textsl{Phys. Rev. Lett.} \textbf{#1}, #3 (#2)}
\newcommand\NPB[3]{\textsl{Nucl. Phys.} \textbf{B#1}, #3 (#2)}

\newcommand\PRD[3]{\textsl{Phys. Rev.} \textbf{D#1}, #3 (#2)}

\newcommand\PLB[3]{\textsl{Phys. Lett.} \textbf{#1B}, #3 (#2)}

\newcommand\AoP[3]{\textsl{Ann. of Phys.} \textbf{#1}, #3 (#2)}

\newcommand\IJMPA[3]{\textsl{Int. J. Mod. Phys.} \textbf{A#1}, #3 (#2)}
\newcommand\IJMPD[3]{\textsl{Int. J. Mod. Phys.} \textbf{D#1}, #3 (#2)}

\begin{document}

\wstoc{For proceedings contributors: Using World Scientific's\\ WS-procs961x669 document class in \LaTeX2e}
{A. B. Author and C. D. Author}

\title{Square-Root Gauge Theory and Modified Gravity --  Gravity-Assisted Confinement/Deconfinement and Emergent Electro-Weak Symmetry Breaking in Cosmology}

\author{Eduardo Guendelman}
\address{Department of Physics, Ben-Gurion University of the Negev, Beer-Sheva, Israel.\\
Frankfurt Institute for Advanced Studies (FIAS),
Ruth-Moufang-Strasse 1, \\
60438 Frankfurt am Main, Germany.\\
Bahamas Advanced Study Institute and Conferences, 
4A Ocean Heights, Hill View Circle, \\
Stella Maris, Long Island, The Bahamas.\\
E-mail: guendel@bgu.ac.il}

\author{Emil Nissimov and Svetlana Pacheva}
\address{Institute for Nuclear Research and Nuclear Energy, Bulgarian Academy of Sciences\\
Boul. Tsarigradsko Chausee 72, BG-1784 ~Sofia, Bulgaria \\
E-mail: nissimov@inrne.bas.bg, svetlana@inrne.bas.bg}

\vspace{-.1in}
\begin{abstract}
In the present contribution to the proceedings of MG17, the main aim is to elucidate the 
physically important effects of a special nonlinear gauge field with a square-root of the 
standard Maxwell Lagrangian in its action, interacting with a specific non-canonical 
modified $f(R)= R + R^2$ gravity formulated 
in terms of metric-independent spacetime volume elements and in addition coupled to the 
bosonic fields of the standard electroweak particle model. When applied in the context of 
cosmological evolution, the above theory consistently describes absence (suppression) of charge 
confinement and electroweak (Higgs) spontaneous breakdown in the ``early'' Universe, 
whereas in the post-inflationary ``late'' universe primarily the presence of the 
``square-root''  nonlinear gauge field dynamically triggers both the appearance of QCD-like 
confinement and dynamical generation of the Higgs effect, as well as dynamical generation of non-zero cosmological constant.

\end{abstract}
\keywords{charge confining square-root nonlinear gauge field system, modified gravity theories, 
non-Riemannian volume-forms; global Weyl-scale symmetry spontaneous breakdown; flat regions 
of scalar inflaton potential}


\bodymatter
\section{Introduction}\label{intro}
Our principal task in the present contribution is to present a consistent approach from first
principles, \textsl{i.e.}, from Lagrangian action principle, to describe consistent
mechanisms driving the appearance, respectively the suppression, of confinement
and electroweak spontaneous symmetry breaking during the various stages in
the cosmological evolution of the Universe \ct{general-cit-1}-\ct{general-cit-7}.

To this effect we recall that the standard model of cosmology is the so called $\L$CDM model,
which came about after the discovery of the late acceleration of the Universe (for review, see \textsl{e.g.} Ref.\ct{accelerateduniverse}. 
Here one thinks of the very ``early'' universe as undergoing an inflationary phase and then asks about what can produce a full history of the universe. A solution for example is the idea of a quintessential  scalar field driving the inflation and then the slowly accelerated phase of the universe as was done in Refs.\ct{quintessentialinflation}.

In what follows we will proceed in two steps. First, we will consider a special kind of a
nonlinear (Abelian or non-Abelian) gauge field theory with a square-root of the
standard Maxwell/Yang-Mills kinetic term which possess the remarkable property of 
producing charge confinement. In what follows it will be called for short ``square-root'' gauge field theory. 

Then we will discuss in some detail the main interesting properties of a new type of 
non-canonical modified (extended) gravity-matter theory, in particular, its implications for  
cosmology. Namely we will consider modified  $f(R)=R+R^2$ gravity coupled in a
non-standard way to a scalar ``inflaton'' field, to the bosonic fields (including the Higgs field) of
the standard electroweak particle model, as well as to the above mentioned ``square-root'' nonlinear gauge field which simulates QCD confining dynamics. Thus, in this way our model will represent qualitatively modified gravity coupled to the whole (bosonic part of the) standard model of elementary particle physics.

The most important non-standard feature of the above model 
is its construction in terms of non-Riemannian spacetime volume-forms
(alternative metric-independent generally covariant volume elements) defined in
terms of auxiliary antisymmetric tensor gauge fields of maximal rank
(see Refs.\ct{susyssb-1,grav-bags} for a consistent geometrical formulation,
which is an extension of the originally proposed method \ct{TMT-orig-1,TMT-orig-2}).
The latter auxiliary volume-form gauge fields were shown in Refs.\ct{grav-bags,grf-essay,AIP-conf}
to be almost {\em pure-gauge} apart from few arbitrary
integration constants, which have the meaning of residual discrete time-conserved degrees of freedom. Thus, they do not produce any additional  {\em propagating field-theoretic}
degrees of freedom (see also Sect.3 below). 

On the other hand the non-Riemannian spacetime volume-forms trigger a series of important
physical features unavailable in ordinary gravity-matter models with the
standard Riemannian volume element (given by the square-root of the determinant
of the Riemannian metric):

(i) The ``inflaton'' $\vp$ develops a remarkable effective scalar potential in
the Einstein frame possessing an infinitely large flat region for large
negative $\vp$ describing the ``early'' universe evolution;

(ii) In the absence of the $SU(2)\times U(1)$ iso-doublet (Higgs) scalar field,
the ``inflaton'' effective potential has another infinitely large flat region
for large positive $\vp$ at much lower energy scale describing the ``late''
post-inflationary (dark energy dominated) universe;

(iii) Inclusion of the $SU(2)\times U(1)$ iso-doublet scalar field $\s$ --
{\em without the usual tachyonic mass and quartic self-interaction term}
-- introduces a drastic change in the total effective scalar potential in the
post-inflationary universe: the effective potential as a function of $\s$
{\em dynamically acquires} exactly the electroweak Higgs-type spontaneous symmetry
breaking form. The latter is a remarkable explicit realization of Bekenstein's idea
\ct{gravity-assist-86} for a gravity-assisted dynamical electroweak spontaneous
symmetry breaking.

(iv) Further important features arise because of the coupling to the above mentioned additional
nonlinear gauge field whose Lagrangian contains a square-root of
the standard Maxwell/Yang-Mills kinetic term. The latter is known to
describe charge confinement in flat spacetime \ct{GG-2} as well as
in curved spacetime for static spherically symmetric field configurations
(Refs.\ct{grav-bags,AIP-conf}; see also Eq.\rf{cornell-type} below).
This is a simple implementation of `t Hooft's idea \ct{thooft} about confinement
being produced due to the presence in the energy density of electrostatic
field  configurations of a term {\em linear} w.r.t. electric displacement field
in the infrared region (arising presumably as an appropriate infrared counterterm).
Therefore, the addition of the ``square-root'' nonlinear gauge field will simulate
the strong interactions QCD-like dynamics.


As a result, in the Einstein frame of the present modified (extended) gravity+matter model the above outlined formalism allows us to achieve:

(a) Bekenstein-inspired \ct{gravity-assist-86} gravity-inflaton-assisted dynamical generation 
of Higgs-type electroweak spontaneous symmetry breaking in the ``late'' universe, while there
is no electroweak breaking in the ``early'' universe;

(b) Simultaneously we obtain gravity-inflaton-assisted dynamical generation of charge
confinement in the ``late'' universe as well as gravity-suppression of confinement,
\textsl{i.e.}, deconfinement in the ``early'' universe.

Finally, in the last Section we will briegly describe a closely related model of modified gravity interacting with a nonlinear ``square-root'' gauge field which describes gravitational bags resembling the solitonic ``constituent quark'' model \ct{const-quark} and MIT bags in QCD phenomenology \ct{MIT-bag-1,MIT-bag-2}.

\section{Charge-Confining Nonlinear Gauge Field}\label{square-root}
We start by first exhibiting the charge-confining feature of a special nonlinear gauge field theory whose Lagrangian action contains an additional term being a square-root of the usual Maxwell Lagrangian.
To this end we will follow the steps of the derivation in Ref.~\refcite{GG-2} of
effective ``Cornell''-type confining potential \ct{cornell-1,cornell-2,cornell-3} between 
quantized charged fermions based on the general formalism \ct{bunster-77} for quantization 
within the canonical Hamiltonian approach a'la Dirac of truncated gauge and gravity theories 
by imposing explicitly spherical symmetry on the pertinent Lagrangian action. 

The corresponding nonlinear gauge field action in curved space-time background with metric 
$g_{\m\n}$ coupled to an external (charged matter) current $J^\m$ reads:
\be
S = \int d^4 x\,\sqrt{-g} \Bigl\lb L(F^2) + A_\m J^\m \Bigr\rb \quad, \quad
L(F^2) = - \frac{1}{4} F^2 - \frac{f_0}{2}\sqrt{-F^2} 
\lab{NL-action}
\ee
with dimensionful coupling constant $f_0$ of the nonlinear gauge field term and where: 
\be
F^2 \equiv F^2(g) = F_{\m\n} F_{\k\l} g^{\m\k} g^{\n\l} \quad ,\quad
F_{\m\n} = \pa_\m A_\n - \pa_\n A_\m \; .
\lab{Fabel-def}
\ee
The action \rf{Fabel-def} yields the following equations of motion:
\be
\pa_\n \Bigl(\sqrt{-g}4 L^\pr (F^2) F^{\m\n}\Bigr) + \sqrt{-g} J^\m = 0
\quad ,\quad L^\pr (F^2) = -\frac{1}{4}\Bigl( 1 - \frac{f_0}{\sqrt{-F^2}}\Bigr) \; ,
\lab{NL-eqs}
\ee
whose $\m=0$ component -- the nonlinear ``Gauss law'' constraint equation reads:
\be
\frac{1}{\sqrt{-g}} \pa_i \bigl(\sqrt{-g} D^i\bigr) = J^0 \quad ,\quad 
D^i = \Bigl( 1 - \frac{f_0}{\sqrt{-F^2}}\Bigr) F^{0i} \; ,
\lab{NL-gauss-law}
\ee
with $\vec{D}\equiv (D^i)$ denoting the electric displacement field
nonlinearly related to the electric field $\vec{E}\equiv (F^{0i})$ as in the
last relation \rf{NL-gauss-law}. 

The special nonlinear gauge field theory \rf{NL-action} possesses a nontrivial vacuum solution 
$\sqrt{-F^2_{\rm vac}} = f_0$, which according to the second Eq.\rf{NL-gauss-law} implies simultaneously:
(a) vanishing of the electric displacement field, $\vec{D}=0$ meaning zero observed charge, and at the same time (b) 
a nontrivial electric field $\vec{E}$. This can be viewed as the simplest classical manifestation of charge 
confinement: $\vec{D}=0$ and nontrivial $\vec{E}$. 

Indeed, for instance in static spherically symmetric fields in a static spherically symmetric space-time metric, \textsl{e.g.} of the form:
\be
ds^2 = g_{\m\n} dx^\m dx^\n = - \cA (r) dt^2 + \frac{dr^2}{\cA (r)}
+ r^2 \bigl( d\th^2 + \sin^2 \th d\phi\Bigr) \; ,
\lab{deSitter-type} 
\ee
with general $\cA (r) = - g_{00}=1/g_{rr}$, the only surviving component of $F_{\m\n}$ is 
the non-vanishing radial component of the electric field 
$E^r = - F_{0r}$, so that $\sqrt{-F^2_{\rm vac}} = \sqrt{2} |\vec{E}|= f_0$. 

In order to exhibit the charge-confining feature of the nonlinear gauge theory \rf{NL-action}
we will employ the canonical Hamiltonian treatment in Ref.~\refcite{bunster-77} 
and will truncate the non-linear gauge field action to purely spherically
symmetric fields, \textsl{i.e.}, we will take $F_{0r} = \pa_0 A_r - \pa_r A_0$ independent
of the space angles and the rest of the components of $F_{\m\n}$ being zero and also will use the specific example of  
a spherically symmetric de Sitter-type metric (cf. \rf{deSitter-type}. 
The action of the truncated theory reads:
\be
S_{\rm truncated} = \int dt \int dr 4\pi r^2 \Bigl\lb \h F_{0r}^2 - 
\frac{f_0}{\sqrt{2}}|F_{0r}| + A_0 J^0 + A_r J^r \Bigr\rb \; .
\lab{NL-action-truncated}
\ee
Let us note that in \rf{NL-action-truncated} there is no explicit dependence on the
Riemannian metric coefficient $\cA (r)$ (cf.\rf{deSitter-type}).
It is now straightforward to apply the canonical Hamiltonian quantization
procedure to \rf{NL-action-truncated} within the Dirac formalism for constrained dynamical
systems (\textsl{e.g.} Ref.~\ct{henneaux-bunster}). Obviously, in the
case of de Sitter spacetime the radial coordinate $r$ must be restricted to
vary up to the de Sitter horizon radius $r_H$.

The canonically conjugated momenta w.r.t. $A_0$ and $A_r$ read:
\be
\Pi^0 = 0 \quad ,\quad \Pi^r = 4\pi r^2 \Bigl( F_{0r} - \frac{f_0}{\sqrt{2}}\Bigr) \; ,
\lab{canon-momenta}
\ee
where the first one $\Pi^0 = 0$ is the standard primary Dirac constraint known in any
gauge theory of Yang-Mills type. For the density of the canonical Hamiltonian one obtains:
\be
\cH = \frac{1}{8\pi r^2} \(\Pi^r\)^2 + \frac{f_0}{\sqrt{2}}\Pi^r +
\pi r^2 f_0^2 - A_r J^r + \Pi^r \pa_r A_0 - J^0  A_0 \; .
\lab{canon-H-0}
\ee
Henceforth, for simplicity we will consider the case with no matter current $J^r = 0$.
Time-preservation of the primary constraint $\Pi^0 = 0$, \textsl{i.e.},
$\frac{d}{dt}\Pi^0 = \Bigl\{ \Pi^0 ,\cH \Bigr\}_{\rm PB} = 0$ yields the
standard secondary Dirac constraint -- the ``Gauss law'' constraint:
\be
\Phi_1 (r) \equiv \pa_r \Pi^r + J^0 = 0 \; .
\lab{gauss-law}
\ee

Thus, one has to Dirac-canonically quantize the theory with canonical
Hamiltonian:
\be
H = \int dr \Bigl\lb \frac{1}{8\pi r^2} \bigl(\Pi^r\bigr)^2 + \frac{f_0}{\sqrt{2}}\Pi^r +
\pi r^2 f_0^2 \Bigr\rb
\lab{canon-H-1}
\ee
and with two first-class a'la Dirac constraints $\Phi_{0,1}=0$ ($\Phi_0 \equiv \Pi^0 = 0$ 
and $\Phi_1 = 0$ as in \rf{gauss-law}), which have to be supplemented by two canonically 
conjugate gauge-fixing conditions $\chi_{0,1}$. Since $A_0$ and its
conjugate momentum $\Pi^0 = 0$ do not mix with the rest of the canonical
variables they have no impact on the pertinent {\em Dirac brackets} between
$A_r$ and $\Pi^r$ to be promoted to quantum operator commutators upon quantization.
Thus we only need to pick an appropriate gauge fixing condition for the ``Gauss law'' 
constraint \rf{gauss-law}, which we can choose in the form:
\be
\chi_1 (r) \equiv \int_{C(r)} dz^\l A_\l (z) \; . 
\lab{Psi-1}
\ee
Here $\int_{C(r)}$ is path integral along a spacelike geodesic $x^\l = x^\l (\xi)$ ending at 
the spacetime point with radial coordinate $r$. In particular, for the interior de Sitter
region ($r \leq r_H$) this spacelike geodesic $x^\l (\xi) = \bigl( t(\xi), r(\xi)\bigr)$ can be taken in the form:
\be
t(\xi) = t = {\rm const} \quad ,\quad r(\xi) = r_H \sin (\xi/r_H) \;\; , \;\;
0 \leq \xi \leq \xi_{\rm fin} \leq r_H \frac{\pi}{2} \;\; , \;\; r(\xi_{\rm fin}) = r \; ,
\lab{dS-curve}
\ee
where $\xi$ is the de Sitter proper distance parameter, so that:
\be
\chi_1 (r) \equiv \int_0^r dz A_r (z) \quad,\quad
\Bigl\{\Phi_1 (r),\chi_1 (r^\pr)\Bigr\}_{\rm PB} = \d (r-r^\pr) \; .
\lab{PB-Phi-Psi}
\ee
Note that here and below $\d (r-r^\pr)$ denotes the Dirac delta-function on the half-line
(both $r, r^\pr > 0$).

It is now straightforward to calculate the Dirac bracket between the
canonically conjugate pair given by:
\br
\Bigl\{ A_r (r), \Pi^r (r^\pr)\Bigr\}_{\rm DB} = 
\Bigl\{ A_r (r), \Pi^r (r^\pr)\Bigr\}_{\rm PB} \phantom{aaaa}
\nonu \\
- \int\int dr^{\pr\pr} dr^{\pr\pr\pr}
\Bigl\{ A_r (r), \Phi_1 (r^{\pr\pr})\Bigr\}_{\rm PB}
\Bigl\{\Phi_1 (r^{\pr\pr}),\chi_1 (r^{\pr\pr\pr})\Bigr\}^{-1}_{\rm PB}
\Bigl\{\chi_1 (r^{\pr\pr\pr}),\Pi^r (r^\pr)\Bigr\}_{\rm PB} \; , \phantom{aaaa}
\lab{DB-def}
\er
by using the standard Poisson bracket 
$\Bigl\{ A_r (r), \Pi^r (r^\pr)\Bigr\}_{\rm PB} = \d (r-r^\pr)$, which yields:
\be
\Bigl\{ A_r (r), \Pi^r (r^\pr)\Bigr\}_{\rm DB} = 2 \d (r-r^\pr) \; .
\lab{DB}
\ee
Upon canonical quantization \rf{DB} becomes:
\be
\llb {\widehat\Pi}^r (r), {\widehat A}_r (r^\pr) \rrb = 2i \d (r-r^\pr)
\quad, \;\; {\mathrm i.e.}\;\; {\widehat\Pi}^r (r) = -2i\d/\d A_r (r) \; .
\lab{Dirac-CCR}
\ee

Now, following Ref.~\refcite{GG-2} we consider a gauge invariant quantum state of two
oppositely charged ($\pm e_0$) fermions located at $r=0$ and $r=L$, respectively, which is explicitly given by:
\be
\left | \Phi \rrangle \equiv \left | {\bar\Psi}(L) \Psi (0) \rrangle =
{\bar\Psi}(L) \exp\Bigl\{ie_0 \int_0^L dz A_r (z)\Bigr\}\Psi (0) \left | 0 \rrangle
\; .
\lab{two-fermion-state}
\ee
The average of the quantized canonical Hamiltonian \rf{canon-H-1} in this state 
\rf{two-fermion-state}: 
\be
\langle \Phi \vert {\widehat H} \vert \Phi\rangle 
\equiv V_{\rm eff} (L)
\lab{V-eff}
\ee
can be viewed as effective potential between the quantized fermionic pair generated by 
the nonlinear gauge field theory containing the ``square-root'' Maxwell term
\rf{NL-action}.

Using relation \rf{Dirac-CCR} one calculates:
\br
\Bigl\lb {\widehat\Pi}^r (r) , ie_0 \int_0^L dz A_r (z) \Bigr\rb = 
2 e_0 \th (L-r) \; , 
\lab{commute-1} \\
\Bigl\lb \Bigl\lb \({\widehat\Pi}^r (r)\)^2 , ie_0 \int_0^L dz A_r (z) \Bigr\rb ,
ie_0 \int_0^L dz A_r (z) \Bigr\rb = 8 e^2_0 \th (L-r) \; , 
\lab{commute-2}
\er
where $\th (r-r^\pr)$ denotes the step-function on the half-line (both $r, r^\pr > 0$).
Upon plugging \rf{commute-1}-\rf{commute-2} into \rf{V-eff} we obtain for the effective potential \rf{V-eff}:
\be
V_{\rm eff} (L) = - \frac{e_0^2}{2\pi} \frac{1}{L} + e_0 f_0 \sqrt{2}\, L +
\bigl( L{\rm -independent} ~{\rm const} \bigr) \; ,
\lab{cornell-type}
\ee
which has precisely the form of the ``Cornell'' potential
\ct{cornell-1,cornell-2,cornell-3}, \textsl{i.e.}, a sum of two pieces --  
an ordinary Coulomb plus a {\em linear confining} term.

In fact, we could equally well take the ``square-root'' nonlinear gauge field 
$A_\m$ to be non-Abelian. We notice that for static spherically symmetric solutions 
the non-Abelian model effectively reduces to the abelian one \ct{GG-2}.
Thus, the ``square-root'' gauge field will {\em simulate the QCD-like confining dynamics}.
In this sense the constants $f_0$ and $e_0$ in \rf{cornell-type} will play the role of a confinement-strength coupling  constant and of a ``color'' charge, respectively. It is with this interpretation that we will view the nonlinear gauge field action \rf{NL-action} in Section 4.

\section{Non-Riemannian Volume-Forms - Basic Properties}
\label{nonrriemannian-gen}
Since non-Riemannian volume-forms, or metric-independent space-time volume elements (equivalently, generally-covariant integration measures on the space-time manifold) will play a fundamental role in formulating our specific version of modified (extended) gravitational theory, here we will briefly describe basic features.

Volume-forms (space-time volume elements) are given by nonsingular maximal rank differential forms 
$\om$:
\br
\int_{\cM} \om \bigl(\ldots\bigr) = \int_{\cM} dx^D\, \Omega \bigl(\ldots\bigr)
\;\; ,\;\;
\om = \frac{1}{D!}\om_{\m_1 \ldots \m_D} dx^{\m_1}\wedge \ldots \wedge dx^{\m_D}\; ,
\lab{omega-1} \\
\om_{\m_1 \ldots \m_D} = - \vareps_{\m_1 \ldots \m_D} \Omega \;\; ,\;\;
dx^{\m_1}\wedge \ldots \wedge dx^{\m_D} = \vareps^{\m_1 \ldots \m_D}\,  dx^D \; .
\lab{omega-3}
\er
Here we are using the following conventions for the alternating symbols $\vareps^{\m_1,\ldots,\m_D}$
 and $\vareps_{\m_1,\ldots,\m_D}$: $\vareps^{01\ldots D-1}=1$ and $\vareps_{01\ldots D-1}=-1$).
The volume element (integration measure density) $\Omega$ transforms as scalar
density under general coordinate reparametrizations.

In standard generally-covariant theories (in $D$ space-time dimensions with Lagranian action 
$S=\int d^D\! x \sqrt{-g} \cL$) the usual Riemannian spacetime volume-form is defined through 
the ``D-bein'' (frame-bundle) canonical one-forms $e^A = e^A_\m dx^\m$ ($A=0,\ldots ,D-1$):
\br
\om = e^0 \wedge \ldots \wedge e^{D-1} = \det\Vert e^A_\m \Vert\,
dx^{\m_1}\wedge \ldots \wedge dx^{\m_D} \;\; ,    
\nonu \\
\longrightarrow \quad
\Omega = \det\Vert e^A_\m \Vert\, d^D x = \sqrt{-\det\Vert g_{\m\n}\Vert}\, d^D x \; .
\lab{omega-riemannian}
\er

In fact, in order to define a manifestly generally-invariant action principle there is 
{\em no a priori} any obstacle to employ instead of $\sqrt{-g}$ another
alternative {\em non-Riemannian} volume element as in \rf{omega-1}-\rf{omega-3}
given by a non-singular {\em exact} $D$-form $\om = d B$ where:
\be
B = \frac{1}{(D-1)!} B_{\m_1\ldots\m_{D-1}}
dx^{\m_1}\wedge\ldots\wedge dx^{\m_{-1}} \; ,
\lab{B-form}
\ee
so that the {\em non-Riemannian} volume element reads:
\be
\Omega \equiv \Phi(B) =
\frac{1}{(D-1)!}\vareps^{\m_1\ldots\m_D}\, \pa_{\m_1} B_{\m_2\ldots\m_D} \; .
\lab{Phi-D}
\ee

Here $B_{\m_1\ldots\m_{D-1}}$ is an auxiliary rank $(D-1)$ antisymmetric tensor gauge
field. $\Phi(B)$, which is in fact the density of the dual of the rank $D$ field strength
$F_{\m_1 \ldots \m_D} = \frac{1}{(D-1)!} \pa_{\lb\m_1} B_{\m_2\ldots\m_D\rb}
= - \vareps_{\m_1 \ldots \m_D} \P (B)$,
similarly transforms as scalar density under general coordinate reparametrizations.

At this point it is crucial to emphazise that the presence of non-Riemannian volume element 
$\Phi(B)$ in a gravity-matter action $S=\int d^D\! x \Phi(B) \cL + \ldots$ {\em does not} 
change the number of {\em field-theoretic} degrees of freedom -- the latter remains the same as 
with the standard Riemannian measure $\sqrt{-g}$. This important fact was demonstrated in detail 
by systematic application (see \textsl{e.g.} \ct{AIP-conf}) of the canonical Hamiltonian 
analysis which reveals that 
the auxiliary tensor gauge field  $B_{\m_1\ldots\m_{D-1}}$ is (almost) {\em pure-gauge}! 
This is because the total Lagrangian is only linear w.r.t. $B$-velocities, 
so it leads to Hamiltonian constraints a'la Dirac.
The only remnant of $B_{\m_1\ldots\m_{D-1}}$ is a {\em discrete degree of freedom}
which appears as integration constant $M$ in the equations of motion w.r.t.
$B_{\m_1\ldots\m_{D-1}}$ (see subsect. 4.2 below). $M$ is in fact a conserved Dirac 
constrained canonical momentum conjugated to the ``magnetic'' $B$-component
$\frac{1}{(D-1)!}\vareps^{i_1\ldots i_{D-1}} B_{i_1\ldots i_{D-1}}$.

\section{Confining Nonlinear Gauge Field Coupled to Modified Gravity and the Bosonic Sector of the Electroweak Standard Model}
\label{noncanon-grav}
Now we proceed to our main task to consider in some detail a specific non-canonical modified gravity theory coupled to the bosonic sector of the electroweak standard model and to the nonlinear ``square-root'' confining gauge field \rf{NL-action}, as well as to exhibit the impact of the latter on the process of emergence of some fundamental processes during evolution of the Universe, notably charge confinement and Higgs mechanism of electroweak symmetry breaking.
\subsection{Modified $f(R)$-Gravity Model with Non-Riemannian Spacetime
Volume-Forms}
\label{TMMT}
We start with the following non-canonical $f(R)=R+R^2$ gravity-matter action 
constructed in terms of two different non-Riemannian volume-forms, \textsl{i.e.} generally
covariant metric-independent volume elements  
(for simplicity we use units with the Newton constant $G_N = 1/16\pi$):
\br
S = \int d^4 x\,\P (A) \Bigl\lb R + L_1 (\vp,X) + L_2 (\s,Y) 
- \h f_0 \sqrt{-F^2}\Bigr\rb +
\nonu \\
\int d^4 x\,\P (B) \Bigl\lb \eps R^2 - \frac{1}{4e^2} F^2  
- \frac{1}{4g^2} \cF^2(\cA) - \frac{1}{4g^{\pr\,2}} \cF^2(\cB) +
\frac{\P (H)}{\sqrt{-g}}\Bigr\rb  \; .
\lab{TMMT-1}
\er
Here the following notations are used:
\begin{itemize}
\item
$\P(A)$ and $\P (B)$ are two independent non-Riemannian volume-forms given in 
terms of the dual field-strengths of two auxiliary rank 3 antisymmetric tensor gauge fields 
$A_{\n\k\l}$ and $B_{\n\k\l}$.
\be
\P (A) = \frac{1}{3!}\vareps^{\m\n\k\l} \pa_\m A_{\n\k\l} \quad ,\quad
\P (B) = \frac{1}{3!}\vareps^{\m\n\k\l} \pa_\m B_{\n\k\l} \; .
\lab{Phi-1-2}
\ee
\item
$\P (H)$ is the dual field-strength of an additional auxiliary tensor gauge field 
$H_{\n\k\l}$, whose presence is crucial for the consistency of the model 
\rf{TMMT-1}:
\be
\P (H) = \frac{1}{3!}\vareps^{\m\n\k\l} \pa_\m H_{\n\k\l} \; .
\lab{Phi-H}
\ee
\item
Let us particularly emphasize that we start within the first-order 
{\em Palatini formalism} for the scalar curvature $R$ and the Ricci tensor
$R_{\m\n}$: $R=g^{\m\n} R_{\m\n}(\G)$, where $g_{\m\n}$, $\G^\l_{\m\n}$ --
the metric and affine connection are {\em apriori} independent.
\item
$L_1 (\vp,X)$ is the scalar ``inflaton'' Lagrangian:
\be
L_1 (\vp,X) = X - f_1 e^{-\a\vp} \;\; ,\;\;
X \equiv - \h g^{\m\n} \pa_\m \vp \pa_\n \vp \; ,
\lab{L-1}
\ee
where $\a, f_1$ are dimensionful positive parameters.
\item
$\s \equiv (\s_a)$ is a complex $SU(2)\times U(1)$ iso-doublet Higgs-like 
scalar field with Lagrangian:
\be
L_2 (\s,Y) = Y - m_0^2 \s^{*}_a \s_a \;\; ,\;\; 
Y \equiv - g^{\m\n}(\nabla_\m \s)^{*}_a \nabla_\n \s_a \; ,
\lab{L-2}
\ee
where the gauge-covariant derivative acting on $\s$ reads:
\be
\nabla_\m \s = 
\Bigl(\pa_\m - \frac{i}{2} \t_A \cA_\m^A - \frac{i}{2} \cB_\m \Bigr)\s \; ,
\lab{cov-der}
\ee
with $\h \t_A$ ($\t_A$ -- Pauli matrices, $A=1,2,3$) indicating the $SU(2)$ 
generators and $\cA_\m^A$ ($A=1,2,3$)
and $\cB_\m$ denoting the corresponding electroweak $SU(2)$ and $U(1)$ gauge fields.
\item
The electroweak gauge field kinetic terms are of the standard Yang-Mills form 
(all $SU(2)$ indices $A,B,C = (1,2,3)$):
\br
\cF^2(\cA) \equiv \cF^A_{\m\n} (\cA) \cF^A_{\k\l} (\cA) g^{\m\k} g^{\n\l} \;\; ,\;\;
\cF^2(\cB) \equiv \cF_{\m\n} (\cB) \cF_{\k\l} (\cB) g^{\m\k} g^{\n\l} \; ,
\lab{F2-def} \\
\cF^A_{\m\n} (\cA) = 
\pa_\m \cA^A_\n - \pa_\n \cA^A_\m + \eps^{ABC} \cA^B_\m \cA^C_\n \;\; ,\;\;
\cF_{\m\n} (\cB) = \pa_\m \cB_\n - \pa_\n \cB_\m \; .
\lab{F-def}
\er
\item
Finally, we stress on the additional coupling in the action \rf{TMMT-1} to the ``square-root''
nonlinear (Abelian) gauge field $A_\m$ as in \rf{NL-action}. Since as discussed in the presious Section the latter simulates QCD-like confinement dynamics, we can view \rf{TMMT-1} as a theory describing  modified gravitation interacting with the fields of the whole standard model of elementary particles.
\end{itemize}

Let us note that the structure of action \rf{TMMT-1} is uniquely fixed by the 
requirement for invariance (with the exception of the regular mass term 
of the iso-doublet Higgs-like scalar $\s_a$
under the following global Weyl-scale transformations:
\br
g_{\m\n} \to \l g_{\m\n} \;\; ,\;\; \vp \to \vp + \frac{1}{\a}\ln \l \;\;,\;\; 
A_{\m\n\k} \to \l A_{\m\n\k} \;\; ,\;\; B_{\m\n\k} \to \l^2 B_{\m\n\k} \; ,
\lab{scale-transf} \\
\G^\m_{\n\l} \;,\; H_{\m\n\k} \;,\; \s_a \; ,\; A_\m \; ,\;\cA^A_\m \; ,\; \cB_\m \;
- \; {\rm inert} \; .
\nonu
\er
In fact, as shown in \ct{AIP-conf} one can, instead with the usual Higgs-like mass term, start 
by temporarily introducing an additional auxiliary scalar field $\psi$ with a fully Weyl-scale 
invariant action
$\int d^4 x\llb \P(A) \(-\h g^{\m\n}\pa_\m \psi \pa_\n \psi\) - \psi^2 \s^{*}_a \s_a \rrb$
- $\int d^4 x \P(C) \psi^2$, where $\P(C)$ is another auxiliary (pure-gauge) 
metric-independent volume element of the same form as in \rf{Phi-1-2}. 
Equations of motion w.r.t. $C_{\n\k\l}$ imply on-shell $\psi = m_0 \equiv {\rm const}$, thus recovering \rf{TMMT-1}.

It is also very important to stress that, as demonstrated in Ref.\ct{AIP-conf} where 
systematic canonical Hamiltonian treatment of gravity-matter theories with non-Riemannian
volume-forms has been worked out, all auxiliary tensor gauge fields 
$A_{\m\n\k}, B_{\m\n\k}, H_{\m\n\k}$
are almost {\em pure-gauge} up to few residual discrete degrees of freedom (the dynamically energing integration constants $M_1, M_2,\chi_2$ in the next Subsection).

\subsection{Derivation of the Einstein-Frame Action. Effective Scalar Field Potential}
\label{einstein-frame-derivation}
To this end now we first consider the solutions of the equations of motion of the initial action 
\rf{TMMT-1} which includes the square root term for the gauge fields, and an electroweak structure similar to that of the standard model w.r.t. auxiliary tensor gauge fields $A_{\m\n\l}$, $B_{\m\n\l}$ and $H_{\m\n\l}$.
which acquire the form of the following algebraic constraints:
\br
R + L_1 (\vp,X) + L_2 (\s,Y) -\h f_0 \sqrt{-F^2} = - M_1 = {\rm const} \; ,
\lab{integr-const-1} \\
\eps R^2 - \frac{1}{4e^2} F^2 - \frac{1}{4g^2} \cF^2(\cA) 
- \frac{1}{4g^{\pr\,2}} \cF^2(\cB) + \frac{\P (H)}{\sqrt{-g}}
= - M_2  = {\rm const} \; ,
\lab{integr-const-2}\\
\frac{\P(B)}{\sqrt{-g}} \equiv \chi_2 = {\rm const} \; ,
\lab{integr-const-3}
\er
where $M_1$ and $M_2$ are arbitrary dimensionful and $\chi_2$
arbitrary dimensionless {\em integration constants}. The algebraic constraint
Eqs.\rf{integr-const-1}-\rf{integr-const-3} are the Lagrangian-formalism counterparts of the
Dirac first-class Hamiltonian constraints on the auxiliary tensor gauge fields
$A_{\m\n\l},\, B_{\m\n\l},\, H_{\m\n\l}$ 
as discussed in \ct{AIP-conf} and also in \ct{grav-bags,grf-essay}.

The equations of motion of \rf{TMMT-1} w.r.t. affine connection $\G^\m_{\n\l}$ 
(recall -- we are using Palatini formalism):
\be
\int d^4\,x\,\sqrt{-g} g^{\m\n} \Bigl(\frac{\P_1}{\sqrt{-g}} +
2\eps\,\frac{\P_2}{\sqrt{-g}}\, R\Bigr) \(\nabla_\k \d\G^\k_{\m\n}
- \nabla_\m \d\G^\k_{\k\n}\) = 0 
\lab{var-G}
\ee
yield a solution for $\G^\m_{\n\l}$ as a Levi-Civita connection:
\be
\G^\m_{\n\l} = \G^\m_{\n\l}({\bar g}) = 
\h {\bar g}^{\m\k}\(\pa_\n {\bar g}_{\l\k} + \pa_\l {\bar g}_{\n\k} 
- \pa_\k {\bar g}_{\n\l}\) \; ,
\lab{G-eq}
\ee
w.r.t. to the following {\em Weyl-rescaled metric} ${\bar g}_{\m\n}$:
\be
{\bar g}_{\m\n} = \bigl(\chi_1 + 2\eps\chi_2 R\bigr) g_{\m\n} 
\quad , \quad
\chi_1 \equiv \frac{\P_1 (A)}{\sqrt{-g}} \; ,
\lab{bar-g}
\ee
$\chi_2$ as in \rf{integr-const-3}.
Upon using relation \rf{integr-const-1} and notation \rf{integr-const-3}
Eq.\rf{bar-g} can be written as:
\be
{\bar g}_{\m\n} = \Bigl\lb\chi_1 - 2\eps\chi_2 \Bigl(L_1(\vp,X) + L_2(\s,Y) 
-\h f_0\sqrt{-F^2} + M_1\Bigr)\Bigr\rb g_{\m\n}\; .
\lab{bar-g-1}
\ee

Varying \rf{TMMT-1} w.r.t. the original metric $g_{\m\n}$ and using relations 
\rf{integr-const-1}-\rf{integr-const-3} we have the ``pre-Einstein'' equations:
\be
\chi_1 \Bigl\lb R_{\m\n} + \h\( g_{\m\n}L^{(1)} - T^{(1)}_{\m\n}\)\Bigr\rb -
\h \chi_2 \Bigl\lb T^{(2)}_{\m\n} + g_{\m\n} \(\eps R^2 + M_2\)
- 4\eps R\,R_{\m\n}\Bigr\rb = 0 \; ,
\lab{pre-einstein-eqs}
\ee
with $\chi_1$ and $\chi_2$ as in \rf{bar-g} and \rf{integr-const-3},
and $T^{(1,2)}_{\m\n}$ being the canonical energy-momentum tensors:
\be
T^{(1,2)}_{\m\n} = g_{\m\n} L^{(1,2)} - 2 \partder{}{g^{\m\n}} L^{(1,2)} \; .
\lab{EM-tensor}
\ee
of the initial scalar+gauge field Lagrangians in the original action \rf{TMMT-1}:
\be
L^{(1)} \equiv L_1 (\vp,X) + L_2 (\s,Y) -\h f_0 \sqrt{-F^2} \;\; ,\;\;
L^{(2)} \equiv - \frac{1}{4e^2} F^2 - \frac{1}{4g^2} \cF^2(\cA) 
- \frac{1}{4g^{\pr\,2}} \cF^2(\cB) \; .
\lab{L-1-2-def}
\ee

Taking the trace of Eqs.\rf{pre-einstein-eqs} and using again relation 
\rf{integr-const-1} we solve for the ratio $\chi_1$ \rf{bar-g}:
\be
\chi_1 = 2 \chi_2 \frac{T^{(2)}/4 + M_2}{L^{(1)} - \h T^{(1)} - M_1} \; ,
\lab{chi-1-eq}
\ee
where $T^{(1,2)} = g^{\m\n} T^{(1,2)}_{\m\n}$. Explicitly we obtain from 
\rf{chi-1-eq}:
\be
\chi_1 = \frac{1}{2\chi_2 M_2} \Bigl( f_1 e^{-\a\vp} + m^2_0 \s^{*}\s - M_1 \Bigr)
\lab{chi-1-sol}
\ee

The Weyl-rescaled metric ${\bar g}_{\m\n}$ \rf{bar-g-1} can be written
explicitly as:
\br
{\bar g}_{\m\n} = \chi_1 {\wti\O} g_{\m\n} \;\; ,\;\;
{\wti\O} \equiv \frac{1+\frac{\eps}{M_2}\bigl( 
f_1 e^{-\a\vp} + m^2_0 \s^{*}\s - M_1\bigr)^2}{1+2\eps\chi_2 \bigl({\bar X} +
{\bar Y} - \h f_0 \sqrt{-{\bar F}^2}\bigr)} \; ,
\lab{bar-g-2} \\
{\bar X} \equiv - \h {\bar g}^{\m\n} \pa_\m \vp \pa_\n \vp \;\; ,\;\;
{\bar Y} \equiv - {\bar g}^{\m\n}(\nabla_\m \s)^{*}_a \nabla_\n \s_a \;\;,
\;\; {\bar F}^2 \equiv F_{\m\n}F_{\k\l} {\bar g}^{\m\k} {\bar g}^{\n\l} \; .
\lab{X-Y-F-bar}
\er

Now, we can bring Eqs.\rf{pre-einstein-eqs} into the standard form of Einstein 
equations in the second-order formalism for the Weyl-rescaled  metric
${\bar g}_{\m\n}$ \rf{bar-g-2}, \textsl{i.e.}, the {\em Einstein-frame} equations: 
\be
R_{\m\n}({\bar g}) - \h {\bar g}_{\m\n} R({\bar g}) = \h T^{\rm eff}_{\m\n}
\lab{eff-einstein-eqs}
\ee
with effective energy-momentum tensor corresponding according to the definition 
\rf{EM-tensor}:
\be
T^{\rm eff}_{\m\n} = g_{\m\n} L_{\rm eff} - 2 \partder{}{g^{\m\n}} L_{\rm eff}
\lab{T-eff}
\ee
to the following effective {\em Einstein-frame} matter Lagrangian (using
short-hand notations \rf{L-1-2-def}, and with $\chi_1$ as in \rf{chi-1-sol}
and ${\wti\O}$ as in \rf{bar-g-2}):
\be
L_{\rm eff} = \frac{1}{\chi_1 {\wti \O}}\Bigl\{ L^{(1)} + M_1 +
\frac{\chi_2}{\chi_1 {\wti\O}}\Bigl\lb L^{(2)} + M_2 + 
\eps (L^{(1)} + M_1)^2\Bigr\rb\Bigr\} \; .
\lab{L-eff}
\ee

Thus, for the the full Einstein-frame action,
\footnotemark\footnotetext{In the absence of the ``square-root'' gauge field, the structure of the 
Einstein-Frame Action and the associated  effective scalar field potential with applications to quintessential 
inflationary scenarios was studied in Refs.\ct{quintinMMT}.} 
where all quantities defined w.r.t.
Einstein-frame metric \rf{bar-g} are indicated by an upper bar, we obtain:
\be
S = \int d^4 x \sqrt{-{\bar g}} \Bigl\lb R({\bar g}) +
L_{\rm eff} \bigl(\vp,{\bar X};\s,{\bar Y}; {\bar F}^2, 
{\bar\cF(\cA)}^2,{\bar\cF(\cB)}^2\bigr)\Bigr\rb \; .
\lab{TMMT-einstein-frame}
\ee
where after using the expressions for $\chi_1$ \rf{chi-1-sol} and $\Omega$ (second Eq.\rf{bar-g-2})  
the explicit form of $L_{\rm eff}$ \rf{L-eff} reads:
\br
L_{\rm eff} = \bigl({\bar X}+{\bar Y}\bigr)\bigl(1-4\eps\chi_2 \cU(\vp,\s)\bigr) +
\eps\chi_2 \bigl({\bar X}+{\bar Y}\bigr)^2 \bigl(1-4\eps\chi_2 \cU(\vp,\s)\bigr)
\nonu \\
- \bigl({\bar X}+{\bar Y}\bigr) \sqrt{-{\bar F}^2} \eps\chi_2\, f_{\rm eff}(\vp,\s) 
- \h f_{\rm eff}(\vp,\s) \sqrt{-{\bar F}^2}
\nonu \\
- \cU(\vp,\s) - \frac{1}{4 e^2_{\rm eff}(\vp,\s)}{\bar F}^2 
-\frac{\chi_2}{4g^2}{\bar\cF}^2(\cA) -\frac{\chi_2}{4g^{\pr\,2}}{\bar\cF}^2(\cB)
\; 
\lab{L-eff-total}
\er
Here ${\bar X},\, {\bar Y},\, {\bar F}^2$ are as in \rf{X-Y-F-bar} (and
similarly for ${\bar\cF(\cA)}^2,\, {\bar\cF(\cB)}^2$).
In \rf{L-eff-total} the following notations are used:
\begin{itemize}
\item
$\cU(\vp,\s)$ is the effective scalar field (``inflaton'' + Higgs-like) potential:
\be
\cU(\vp,\s) = \frac{\bigl( f_1 e^{-\a\vp} + m^2_0 \s^{*}\s - M_1\bigr)^2}{4\chi_2
\bigl\lb M_2 + \eps \bigl( f_1 e^{-\a\vp} + m^2_0 \s^{*}\s - M_1\bigr)^2\bigr\rb} \; .
\lab{U-vp-s}
\ee
\item
$f_{\rm eff}(\vp,\s)$ is the effective confinement-strength coupling constant:
\be
f_{\rm eff}(\vp,\s) = f_0 \bigl(1-4\eps\chi_2 \cU(\vp,\s)\bigr) \; ;
\lab{f-eff}
\ee
\item
$e^2_{\rm eff}(\vp,\s)$ is the effective ``color'' charge squared:
\be
e^2_{\rm eff}(\vp,\s) = \frac{e^2}{\chi_2}
\Bigl\lb 1 + \eps e^2 f_0^2 \bigl(1-4\eps\chi_2 \cU(\vp,\s)\bigr) \Bigr\rb^{-1}
\lab{e-eff}
\ee
\end{itemize}

Note that \rf{L-eff-total} is of quadratic {\em ``k-essence''} type
\ct{k-essence-1,k-essence-2,k-essence-3,k-essence-4} w.r.t. the
``inflaton'' $\vp$ and the Higgs-like $\s$ fields.

\section{Gravity Assisted Confinement/Deconfinement and Emergent Higgs Mechanism during Cosmological Evolution}
\label{quintess}
From the explicit form of $L_{\rm eff}$ \rf{L-eff-total} we find that the nonlinear ``confining'' gauge field $A_\m$ develops a nontrivial vacuum field-strength:
\be
\frac{\pa L_{\rm eff}}{\pa {\bar F}^2}\bgv_{{\bar X},{\bar Y}=0} = 0
\lab{F-vac-eq}
\ee
explicitly given by:
\be
\sqrt{-{\bar F}^2}_{\rm vac} = f_{\rm eff}(\vp,\s)\, e^2_{\rm eff}(\vp,\s)
\lab{F-vac}
\ee
Substituting \rf{F-vac} into \rf{L-eff-total} we obtain the following total
effective scalar field potential (with $\cU(\vp,\s)$ as in \rf{U-vp-s}):
\be
\cU_{\rm total}(\vp,\s) = \frac{\cU(\vp,\s)(1-\eps e^2 f_0^2) + e^2 f_0^2/4\chi_2}{1
+ \eps e^2 f_0^2 \bigl(1-4\eps\chi_2 \cU(\vp,\s)\bigr)} \;\; ,\;\; 
\cU(\vp,\s) ~{\rm as ~in} ~{\rm \rf{U-vp-s}}) \; ,\;
\lab{U-eff-total}
\ee
which has few remarkable properties.

First, $\cU_{\rm total}(\vp,\s)$ \rf{U-eff-total} possesses two infinitely large flat regions
as function of $\vp$ when $\s$ is fixed:

(a) (-) flat ``inflaton'' region for large negative values of $\vp$,

(b) (+) flat ``inflaton'' region for large positive values of $\vp$,

respectively, as graphically depicted on Fig.1 (for $m_0 \s^{*}\s \leq M_1$) or 
Fig.2 (for $m_0 \s^{*}\s \geq M_1$).

\begin{figure}
\begin{center}
\includegraphics[width=4in]{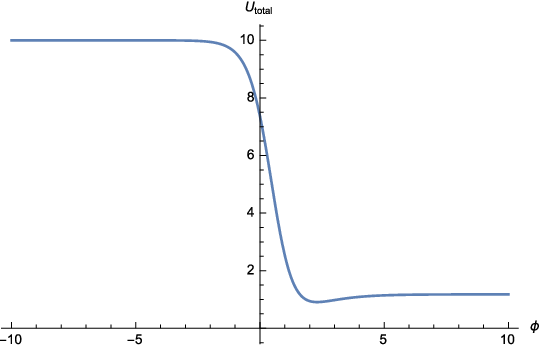}
\caption{Qualitative shape of the total effective scalar potential $U_{\rm total}$ 
\rf{U-eff-total} as function of the ``inflaton''$\vp$ for fixed Higgs-like $\s$
(when $m_0 \s^{*}\s \leq M_1$).}
\end{center}
\end{figure}

\begin{figure}
\begin{center}
\includegraphics[width=4in]{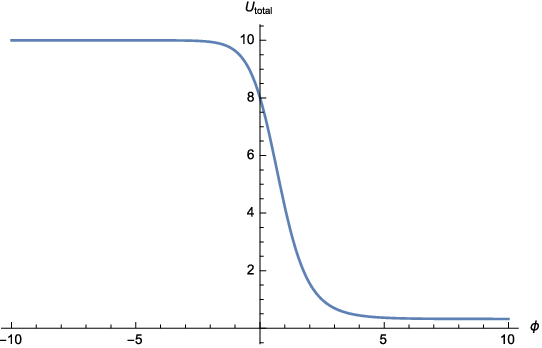}
\caption{Qualitative shape of the total effective scalar potential $U_{\rm total}$ 
\rf{U-eff-total} as function of the ``inflaton''$\vp$ for fixed Higgs-like $\s$
(when $m_0 \s^{*}\s \geq M_1$).}
\end{center}
\end{figure}

\vspace{0.1in}

(i) In the (-) flat ``inflaton'' region:
\begin{itemize}
\item
The effective scalar field potential \rf{U-vp-s} reduces to:
\be
\cU(\vp,\s ={\rm fixed}) \simeq \frac{1}{4\eps\chi_2} \% \quad\longrightarrow\quad
\nonu
\ee
leading for the total scalar potential \rf{U-eff-total} to:
\be
\cU_{\rm total} \simeq \cU^{(-)}_{\rm total} = \frac{1}{4\eps\chi_2} \; ,
\lab{U-minus}
\ee
which implies that all terms containing $\vp$ and $\s$ disappear from the
Einstein-frame Lagrangian \rf{TMMT-einstein-frame}, \textsl{i.e.}, there is
{\em no electroweak spontaneous breakdown in the (-) flat ``inflaton'' region}. 
\item
From \rf{f-eff} the first relation \rf{U-minus} implies $f_{\rm eff} = 0$, 
\textsl{i.e.}, there is {\em no confinement in the (-) flat ``inflaton'' region}, meaning gravity-suppression of confinement in this region.
\end{itemize}

\vspace{0.1in}

(ii) In the (+) flat ``inflaton'' region:
\begin{itemize}
\item
The effective scalar field potential \rf{U-vp-s} becomes:
\be
\cU(\vp,\s) \simeq \cU_{(+)}(\s) = \frac{\bigl(m_0^2 \s^{*}\s - M_1\bigr)^2}{
4\chi_2 \bigl\lb M_2 + \eps \bigl(m_0^2 \s^{*}\s - M_1\bigr)^2\bigr\rb} \; ,
\lab{U-plus} 
\ee
and, accordingly the total scalar potential \rf{U-eff-total} reads:
\be
\cU_{\rm total} (\vp,\s) \simeq \cU^{(+)}_{\rm total}(\s) = 
\frac{\cU_{(+)}(\s)(1-\eps e^2 f_0^2) + e^2 f_0^2/4\chi_2}{1
+ \eps e^2 f_0^2 \bigl(1-4\eps\chi_2 \cU_{(+)}(\s)\bigr)}
\lab{U-total-plus}
\ee
producing a dynamically generated {\em nontrivial vacuum for the Higgs-like field}:
\be
|\s_{\rm vac}|= \sqrt{M_1}/m_0 \; ,
\lab{higgs-vac}
\ee
\textsl{i.e.}, we obtain {\em ``gravity-assisted'' electroweak spontaneous breakdown
in the (+) flat ``inflaton'' region}.
\item
At the Higgs vacuum \rf{higgs-vac} we have dynamically generated vacuum energy density, 
\textsl{i.e.} a {\em dynamically generated cosmological constant} $\L_{(+)}$ which arises 
primarily due to the presence of the ``square-root'' nonlinear gauge field:
\be
\cU^{(+)}_{\rm total}(\s_{\rm vac}) \equiv 2 \L_{(+)} = \eps e^2 f_0^2
\Bigl\lb 4\eps\chi_2 \bigl( 1 + \eps e^2 f_0^2\bigr)\Bigr\rb^{-1} \; .
\lab{CC-plus}
\ee
\item
The effective confinement-strength coupling constant becomes:
\be
f_{\rm eff} \simeq f_{(+)} = f_0 \bigl(1-4\eps\chi_2 \cU_{(+)}(\s)\bigr) > 0\; ,
\lab{f-plus}
\ee
threfore we obtain {\em ``gravity-assisted'' charge confinement in the 
(+) flat ``inflaton'' region}.
\end{itemize}

As seen from Fig.1 or Fig.2, the two flat ``inflaton'' regions of the total scalar field 
potential \rf{U-eff-total} given by $\cU^{(-)}_{\rm total} = \frac{1}{4\eps\chi_2}$ 
\rf{U-minus} and $\cU^{(+)}_{\rm total}(\s_{\rm vac}) \equiv 2 \L_{(+)} = \eps e^2 f_0^2
\Bigl\lb 4\eps\chi_2 \bigl( 1 + \eps e^2 f_0^2\bigr)\Bigr\rb^{-1}$
\rf{CC-plus}, respectively, can be identified (cf. \textsl{e.g.} Refs.\ct{quintessentialinflation}) 
as describing the ``early'' (``inflationary'') and ``late'' (today's dark energy dominated) 
epochs in the cosmological evolution of the universe provided we take the following 
numerical values for the parameters 
in order to conform to the {\em PLANCK} data \ct{Planck-1,Planck-2}:
\be
\cU^{(-)}_{\rm total} \sim 10^{-8} M_{\rm Pl}^4 \to 
\eps\chi_2 \sim 10^8 M_{\rm Pl}^{-4}
\;\; ,\;\; \L_{(+)} \sim 10^{-122} M_{\rm Pl}^4 \to 
\frac{e^2 f_0^2}{\chi_2} \sim 10^{-122} M_{\rm Pl}^4 \; ,
\lab{param-1}
\ee
where $M_{\rm Pl}$ is the Planck mass scale.

From the Higgs v.e.v. $|\s_{\rm vac}|= \sqrt{M_1}/m_0$ and the Higgs mass
$\frac{M_1 m_0^2}{4\chi_2 M_2}$ resulting from the dynamically generated
Higgs-like potential $\cU^{(+)}_{\rm total}(\s)$ \rf{U-total-plus} we find:
\be
m_0 \sim M_{\rm EW} \;\; ,\;\; M_{1,2}\sim M_{\rm EW}^4 \; ,
\lab{param-2}
\ee
where $M_{\rm EW} \sim 10^{-16} M_{\rm Pl}$ is the electroweak mass scale.
\section{``Square-Root'' Nonlinear Gauge Field and Gravitational Bags}
\label{grav-bags}
Here we will very briefly mention another interesting effect produced by the ``square-root'' nonlinear gauge 
field \rf{NL-action} when interacting with special modified gravity similar to \rf{TMMT-1} 
without coupling to the fields of the standard electorweak model ($\s,\cA,\cB$) and with sightly 
more general scalar (called here "dilaton") field Lagrangian than $L_1 (\vp,X)$ \rf{L-1}. The starting action in terms of non-Riemannian volume-elements is:
\be
S = \int d^4 x\,\P_1 (A) \Bigl\lb R + L^{(1)} \Bigr\rb +
\int d^4 x\,\P_2 (B) \Bigl\lb L^{(2)} + \eps R^2 
+ \frac{\P (H)}{\sqrt{-g}}\Bigr\rb \; .
\lab{TMMT+GG}
\ee
where:
\br
L^{(1)} = -\h g^{\m\n} \pa_\m \vp \pa_\n \vp - f_1 \exp \{-\a\vp\} 
- \frac{f_0}{2}\sqrt{- F^2} \; , 
\lab{L-1-GG} \\
L^{(2)} = -\frac{b}{2} e^{-\a\vp} g^{\m\n} \pa_\m \vp \pa_\n \vp + f_2 \exp \{-2\a\vp\}
 - \frac{1}{4e^2} F^2 \; , 
\lab{L-2-GG}
\er
with $b$ a numerical parameter - coupling constant of the non-canonical additional ``dilaton'' kinetic term and where $f_2$ is another dimensionaful ``dilaton'' coupling constant like $f_1$.

Following the same line of derivation as with \rf{L-eff} we arrive at the following Einstein-frame effective matter Lagrangian:
\be
L_{\rm eff} = A(\vp) X + B (\vp) X^2 - {\wti U}_{\rm eff}(\vp)
- \frac{F^2)}{4e_{\rm eff}^2(\vp)}
- \frac{f_{\rm eff}(\vp)}{2}\sqrt{-F^2({\bar g})}
- \eps\chi_2 f_0 A(\vp) X \sqrt{-F^2} \; , 
\lab{L-eff-GG}
\ee
with $X\equiv -\h g^{\m\n} \pa_\m \vp \pa_\n \vp$. The coefficient functions in \rf{L-eff-GG} read:
\be
A(\vp) = 1 - 4 {\wti U}_{\rm eff}(\vp) \Bigl\lb \eps\chi_2  
- \frac{\chi_2 b e^{-\a\vp}}{2(V(\vp)-M_1)} \Bigr\rb \;\;,\;\;
B(\vp) = \eps\chi_2 - 4 {\wti U}_{\rm eff}(\vp) \Bigl\lb \eps\chi_2  
- \frac{\chi_2 b e^{-\a\vp}}{2(V(\vp)-M_1)} \Bigr\rb^2 \; ,
\lab{A-B-GG}
\ee
whereas the effective scalar field potential reads (the same as \rf{U-vp-s} for $f_2=0$ and modulo the Higgs-field terms):
\be
{\wti U}_{\rm eff}(\vp) = 
\frac{\(f_1 e^{-\a\vp}-M_1\)^2}{4\chi_2\,\Bigl\lb 
f_2 e^{-2\a\vp} + M_2 + \eps (f_1 e^{-\a\vp}-M_1)^2\Bigr\rb} \; ,
\lab{U-eff-GG}
\ee
and the effective coupling constants $f_{\rm eff}(\vp), e_{\rm eff}(\vp)$ are of the same form as in
\rf{f-eff}, \rf{e-eff} with $\cU(\vp,\s)$ \rf{U-vp-s} replaced by 
${\wti U}_{\rm eff}(\vp)$ \rf{U-eff-GG}. 

In Ref.\ct{grav-bags} the Einstein-frame effective Lagrangian \rf{L-eff-GG} corresponding to 
model \rf{TMMT+GG} was analyzed in detail with the following findings:

(a) As a function of the ``dilaton'' $\vp$, the effective Lagrangian \rf{L-eff-GG} possesses
two infinite flat regions simialar to Fig.1,2 above: (-) flat region for large negative $\vp$ 
and (+) flat region for large negative $\vp$.

(b) There are two type of ``dilaton'' vacuums: for $\vp = {\rm const}$ (standard vacuum) and 
for $X = {\rm const} $ (``kinetic'' vacuum; here exploiting the quadratic ``k-essence'' form of
\rf{L-eff-GG}). 

As a result we get 3 different types of phases:

(i) For the phase corresponding to the standard vacuum ($\vp = {\rm const}$) 
on both flat $(\pm)$ regions we have confinement since in both $(\pm)$ regions we have 
non-zero effective confinement-strength coupling constant $f_{\rm eff}(\vp)$.

(ii) For the phase corresponding to the ``kinetic'' vacuum $X = {\rm const}$ when $\vp$ is on 
the (+) flat region the total effective confinement-strength coupling constant 
$f_{\rm eff}(\vp)$ vanishes, so there is no confinement. 

(iii) For $X = {\rm const}$ when $\vp$ is on the (-) flat region this once again corresponds to a  charge-confining phase due to non-vanishing $f_{\rm eff}(\vp)$.

As shown in \ct{grav-bags}, in all three cases the spacetime metric we get is
de Sitter or Schwarzschild-de Sitter.
Both ``kinetic vacuums'' (ii) and (iii) can exist only within a
finite-volume space region below a de Sitter horizon. 
Extension to the whole
space requires matching geometry of phase (iii) to a static spherically
symmetric configuration containing the standard constant ``dilaton'' vacuum in the outer region
beyond the de Sitter horizon 
The extension for the geometry of phase (ii) requires more complicated matching 
beyond the de Sitter horizon 
- the exterior region with a nonstandard Reissner-Nordstr{\"o}m-de Sitter geometry carrying 
an additional constant radial background electric field \ct{PLB-2013, belgrad-2013}. 
As a result we obtain two classes of gravitational
bag-like configurations with properties, which on one hand partially parallel some of
the properties of the solitonic ``constituent quark'' model \ct{const-quark} and, on the
other hand, partially mimic some of the properties of MIT bags in QCD phenomenology 
\ct{MIT-bag-1,MIT-bag-2}.

\section{Conclusions}
\label{conclude}
The main focus in the present contribution is on a special kind of 
nonlinear gauge field whose Lagrangian contains a square-root of the usual Maxwell term, 
which is known to describe charge confinement in flat space-time, and study its role and impacts in curved space-time when interacting with specific modified $f(R)=R+R^2$ gravity coupled to the bosonic fields of the standard electroweak model of elementary particles, thus presumably producing notable implications \textsl{e.g.} on cosmological evolution.

There are two sections with preliminary material. In section 2 we review in some detail the derivation of the confining property of the ``square-root'' (Abelian or non-Abelian) gauge cornell
field on quantized point-like fermions in spherically-symmetric curved space-time, generating 
the well-known ``Cornell'' potential \ct{cornell-1,cornell-2,cornell-3} among them. 
In Section 3 we briefly outline the basics of the formalism of the non-Riemannian 
(metric-independent) space-time volume elements, which is the corner stone in the construction 
of broad classes of modified gravitational theories  ......

We then proceed to construct the full action in terms of metric-indepedent volume elements of 
$f(R)=R+R^2$ gravity coupled scalar ``inflaton'' field, to the electroweak bosonic fields and 
to the ``square-root'' nonlinear gauge field which will mimic QCD-like confining dynamics. 
The form of the starting action is uniquely fixed by requiring global Weyl-scale invariance. 
Passing further to the pertinent Einstein-frame action  we obtain a nontrivial total 
effective scalar field potential as well as nontrivial effective ``color'' charge and confinement-strength coupling constant - all functions of the ``inflaton'' $\vp$ and the Higgs $\s$ fields. 

The total effective scalar potential posesses a remarkable property - it has two infinitely large  flat regions as function of $\vp$ ($\s$  fixed): (-) flat region for large negative $\vp$ and 
(+) flat region for large positive $\vp$ (cf. Figs.1,2 above) which are appropriate for description of the evolution of the ``early'' universe (the (-) flat region) and of the ``late'' 
universe (the (+) flat region), accordingly. 

Taking into account the contribution of the nontrivial ``square-root'' gauge field vacuum 
\rf{F-vac} to the total effective scalar potential we find that in the (-) flat region there is 
{\em no} generation of symmetry breaking Higgs potential, as well as the effective 
(field-dependent) confinement-strength effective coupling constant vanishes. Thus, in the ``early'' universe both charge confinement and electroweak symmetry breaking are gravity-suppressed. On the other hand, in the (+) flat region both the confinement-strength effective coupling is non-zero, as well as the standard electroweak symmetry-breaking Higgs potential is dynamically 
generated. On top of this thanks to the nonvanishing confinement-strength coupling of the 
``square-root'' nonlinear gauge field a nontrivial non-zero cosmological constant is 
{\em dynamically} generated responsible for the late-time accelerated expansion of the universe.

In the final Section 6 we briefly touched the interesting phenomenon of dynamical creation of 
gravitational ``bags'' by the ``square-root'' gauge field coupled to modified $f(R)=R+R^2$ 
gravity plus scalar ``dilaton'', which resemble the sonstituent quak model and/or the MIT bags in QCD phenomenology. 

Finally, let us also mention other noteworthy effects of the nonlinear ``square-root'' gauge field 
when coupled to modified gravity-matter (cf.\rf{TMMT-1} without the electroweak bosonic fields).
Namely, as shown in Refs.\ct{PLB-2013,belgrad-2013} the latter produces 
non-standard black hole solutions with constant vacuum electric field
and with ``hedge-hog''-type spacetime asymptotics, which are shown to obey the
first law of black hole thermodynamics; new generalized Levi-Civita-Bertotti-Robinson 
type``tubelike'' space-times; new types of charge-``hiding'' and charge-confining ``thin-shell'' wormhole solutions.

\section*{Acknowledgments}
We gratefully acknowledge support of our collaboration through 
the academic exchange agreement between the Ben-Gurion University in Beer-Sheva,
Israel, and the Bulgarian Academy of Sciences. 
E.G. has received partial support from European COST actions CosmoVerse - COST Action CA21136 and CA23130 Bridging high and low energies in search of quantum gravity (BridgeQG), 
E.N. and S.P. have received partial support from European COST actions CA16104, CA18108. 

\end{document}